\input article.sty\relax
\@addtoreset{equation}{section}
\newcounter{subeqn}

\def\aeqn{\setcounter{subeqn}{1}}
\def\neqn{\addtocounter{equation}{-1}\addtocounter{subeqn}{1}}
\def\zeqn{\setcounter{subeqn}{0}}
\documentstyle[dina4,12pt]{myart}
\def \bx {\mbox{\bf x}}
\def \Hhat {\hat{H}}
\def \that {\hat{t}}
\def \That {\hat{T}}
\def \v {\hat{v}}
\def \Vhat {\hat{V}}
\def \rhat {\hat{\rho}}

\def \A {\hat{A}}
\def \S {\hat{S}}

\newcommand{\ie}{{\em i.e.\/}}

\newcommand{\etal}{{\em et al.\/}}

\begin{document}
\title{Correlation Dynamics of  Green's
Functions\thanks{This work was supported in part
by the National Natural Science Foundation and the Doctoral Education
Fund of the State Education Commission of China, the Director, Office
of Energy Research, Office of High Energy and Nuclear Physics, Nuclear
Physics Division of the U.S.Department of Energy under Contract
DE-AC03-76SF00098, by the Deutsche Forschungsgemeinschaft and GSI Darmstadt.}}
\author{Shun-Jin Wang$^{1,2,3}$,  Wei Zuo$^1$ and Wolfgang Cassing$^2$\\
{\it $^1$ Department of Modern Physics, Lanzhou University}\\
{\it Lanzhou 730000, PR China } \\
{\it $^2$ Institut fr Theoretische Physik, Universitt Giessen,}\\
{\it 6300 Giessen, Germany}\\
{\it $^3$ Nuclear Science Division, Lawrence Berkeley Laboratory,}\\
{\it University of California, Berkeley, CA 94720,USA}}
\maketitle
\begin{abstract}
We generalize the methods used in the theory of correlation dynamics
and establish a set of equations of motion for
many-body correlation green's functions in the non-relativistic
case. These non-linear and coupled equations of motion describe
the dynamical evolution of correlation green's functions of different
order  and transparently show how many-body correlations
are generated by the different interaction terms in a genuine
nonperturbative framework. The nonperturbative results
of the conventional green's function theory are included in the
present formalism as two limiting cases (the so-called ladder
diagram summation and ring diagram summation) as well as  the
familiar  correlation dynamics of density-matrices in the equal time
limit. We present  explicit expressions for three- and
four-body correlation functions that can be used to dynamically restore the
trace relations for spin-symmetric fermi systems and study numerically
the relative importance of two-, three- and four-body correlations for
nuclear configurations close to the groundstate.
\end{abstract}
\newpage
\zeqn
\section{Introduction}
The Green's function approach is a well-known  method in quantum many-body
problems and quantum field theory. In fact, quantum electrodynamics
and quantum field theory are generally formulated in terms of green's
functions \cite{r1}; they are also
widely used in condensed matter physics and nuclear physics \cite{r2}.
With the extension to finite temperature green's functions \cite{r3}
the quantum equilibrium statistical physics has been rapidly developed, too.
Ever since the pioneering work of Schwinger \etal \cite{r4,r5} was published,
 the non-equilibrium green's function theory
 and the closed time path green's function theory
have been widely explored \cite{r6} - \cite{r9}. Nowadays it has
become a powerful tool for describing non-equilibrium quantum
phenomena \cite{r10} - \cite{r15}.

Before the 1960's, the practically tractable green's function theory
was, in fact, a perturbation theory expanded in terms of the interaction
strength \cite{r16}. Equipped with renormalization method, generating
functional, and Feynman diagram technique, the calculation of
perturbative green's functions became simple and transparent.
However, for systems with strong interactions or repulsive hard cores,
the naive perturbation theory does not work and non-perturbative
methods are needed. The most obvious solution is to modify the
conventional perturbation theory and combine the series expansion
with a certain summation rule. Since the late 1950's, the non-perturbative
method based on the conventional green's function was  just
aiming at  that. In order
to derive a set of non-perturbative integro-differential equations based
on some infinite summation rules, one has to resort to the intuition
of Feynman diagrams \cite{r2}.  Apparently, these nonperturbative
approaches  depend strongly on Feynman diagrams and
their selection rules, and hence the completeness, the unification,
and the systematics are lacking for such  approaches.

To remedy the above shortcomings, Martin, Schwinger, and Kadanoff (MSK)
\cite{r4,r5}
developed a new hierarchy which provides a complete set of equations
of motion for many-body Green's functions. This approach is a
multi-time generalization of the BBGKY \cite{r25}
hierarchy from the density-matrix
formalism to the Green's function formalism. This hierarchy allows for a
nonperturbative treatment of the many-body problems, if a truncation
scheme is provided properly. However,
just like the BBGKY hierarchy, the MSK hierarchy does not
provide an explicit treatment of the many-body correlations since the
equations of motion in the MSK formalism include the unlinked many-body
Green's functions $G^{(n)}$, not for the linked many-body correlation
Green's functions $G^{(n)}_c$ (that are formulated in this paper),
and enormous unlinked terms
are present in its equations of motion. Like the BBGKY hierarchy,
there are two similar reasons that cause the
above problems: (i) The explicit cluster expansion is not incorporated
into the Martin-Schwinger's equations of motion systematically to all orders.
(ii) The unlinked terms are not removed from the equations of motion.
Due to the above reasons, just as in case of the BBGKY hierarchy, the MSK
hierarchy is a dynamical theory for the unlinked many-body Green's
functions, not for the linked many-body Green's functions.

It is the purpose of this article to develop a dynamical theory
for the correlated Green's functions within
the framework of the MSK hierarchy.
The theory we aim at is not an expansion in terms of the interaction strength,
but in terms of many-body correlations. Moreover, it should provide
a reasonable truncation scheme by the equations of motion  and increase
in accuracy when including higher order  correlations.
Each truncation scheme will lead to a correlation dynamics
of certain order $n$; each of them is nonlinear and nonperturbative.

In the past the correlation dynamics has been widely explored
\cite{r17} - \cite{r21} and
its applications to heavy-ion physics \cite{r22} - \cite{r24}
have motivated
the research above and also strengthened the desire for developing a
correlation dynamics of green's functions.  As it will turn out,
the correlation density-matrix theory is,
in fact, a special equal time limit of the green's function theory.
For the latter it has
been shown \cite{r17} that the existing nuclear many-body theories,
such as HF, TDHF, ETDHF, HF-Brueckner, second RPA, Faddeev etc.,
can be obtained from
the quantum correlation dynamics within its lowest orders of
truncation.  Moreover, the wide-spread used BUU equation can be
derived in the framework of two-body correlation dynamics
adopting semi-classical
limits within the phase-space representation \cite{r18} - \cite{r21}.
In ref. \cite{r26} it was furthermore shown that the two-body correlation
dynamics in the small amplitude limit provides a more general
framework than the second RPA, which is regained within simple limits.
Realistic calculations based on the
two-body correlation dynamics (TDDM calculations) have  been
performed  for the damping of nuclear giant resonances in
\cite{r27} and for
low energy heavy-ion collisions in \cite{r28}.
In addition, the dynamical equations of motion for the two-body correlation
density matrix have been used in a nonperturbative study of the
damping of giant resonances in hot nuclei \cite{r29} and shown,
that the full two-body correlations, namely the joint
effect of all the ladder diagrams, the ring diagrams and mixed diagrams
are important to explain the long standing phenomenon.
The above results assure that the correlation dynamics
of green's functions will reach similar or, hopefully, even better results.

The article is organized as follows. In sect.2 the many-body
correlation dynamics of green's functions for a non-relativistic
quantum system is formulated in detail.  In sect.3 the normal equal
time limit of the  dynamical equations is taken and the
density-matrix correlation theory is recovered in this limit.
Sect.4 is devoted to a brief reminder on several truncation schemes,
which are related to
 TDHF in the one-body limit, time-dependent G-matrix theory
(incorporating ladder
diagrams or short range correlations) and/or  ring diagrams (long range
correlations) in a unified manner. Since the traditional truncation schemes
violate the fundamental trace relations \cite{r30,r31} the dynamical
restoration of these relations for spin-symmetric systems is discussed, too,
and explicit calculations for the relative strength of two-, three- and
four-body correlations are presented for nuclear configurations close to the
groundstate. In sect.5 we finally give a summary and discussion of open topics.
\newpage

\zeqn
\section{Correlation Dynamics of  Many-Body Green's Functions}
\subsection{\it Equations
of Motion for Field Operators }

We consider an interacting fermion system described in terms of
field operators
\(\psi(\bx,t) \) and \(\psi^\dagger(\bx,t)\).
The fermion field may contain
several components which  represent  different kinds of fermions,
such as leptons  or baryons. For physically
interesting systems, the Hamiltonian can be written as \cite{r2}
\vspace{2 mm}
\begin{equation}
\label{e201}
\Hhat=\int \psi^\dagger(\bx,t)\that(\bx)\psi(\bx,t)d^3\bx
  +{1\over 2}\int\int\psi^\dagger(\bx,t)\psi^\dagger(\bx',t)\v(\bx,\bx')
   \psi(\bx',t)\psi(\bx,t)d^3\bx d^3\bx',
\end{equation}
where $\psi(\bx,t)$ and $\psi^\dagger(\bx,t)$ satisfy anti-commutation
relations,
\aeqn
\begin{equation}
\label{e202}
[\psi(\bx,t),\psi^\dagger(\bx',t) ]_+=\delta^{(3)}(\bx-\bx'),
\end{equation}
\neqn
\begin{equation}
  [\psi(\bx,t),\psi(\bx',t)]_+=[\psi^\dagger(\bx,t),\psi^\dagger
      (\bx',t)]_+= 0.
\end{equation}
\zeqn
\noindent
The time evolution of the field operators then obeys Heisenberg equations,
\vspace{2 mm}
\begin{equation}
\label{e203}
\aeqn
   i\partial_t\psi(\bx,t)=[\psi(\bx,t),\Hhat]
        =\that(\bx)\psi(\bx,t)+\int d^4x'\psi^\dagger(x')\Vhat(x,x')
        \psi(x')\psi(x),
\end{equation}
\vspace{2 mm}
\neqn
\begin{equation}
i\partial_t\psi^\dagger(\bx,t)=[\psi^\dagger(\bx,t),\Hhat]
=-\that(\bx)\psi^\dagger(\bx,t)-\int d^4x'\psi^\dagger(x)
\psi^\dagger(x')\Vhat(x,x')\psi(x').
\end{equation}
\zeqn

\vspace{2 mm}
In eqs.(2.3a,b), the following notation of interactions in 4-dimensional
space-time has been introduced,
\vspace{2 mm}

\begin{equation}
\label{e204}
\Vhat(x,x')=\Vhat(x',x)=\delta(t-t')\v(\bx-\bx').
\end{equation}

\subsection{\it Many-Body Green's Functions and Their Equations of
Motion}

All information about a many-body system can be obtained from its green's
functions which are conventionally defined, for 1-, 2-,... and n-body
green's functions as
\vspace{2 mm}
\begin{equation}
\label{e205}
iG(1;1')=<\That\psi(1)\psi^\dagger(1')>
\end{equation}
\vspace{2 mm}
\begin{equation}
\label{e206}
i^2G^{(2)}(1,2;1',2')=
<\That\psi(1)\psi(2)\psi^\dagger(2')\psi^{\dagger}(1')>
\end{equation}
\vspace{2 mm}
\[\ldots\ldots\]
\vspace{2 mm}
\begin{equation}
\label{e207}
i^nG^{(n)}(1,...n;1',...n')=
<\That\psi(1)...\psi(n)\psi^\dagger(n')...\psi^\dagger(1')>.
\end{equation}
where \(\psi(x_i)\) and \(\psi^\dagger(x_i)\) are field operators in the
Heisenberg picture and $\That$  is a time ordering operator. The average
of a field-operator product is defined as a quantum mechanical and
ensemble average,
\vspace{2 mm}
\begin{equation}
\label{e208}
i^nG^{(n)}(1,...n;1',...n')=
<......>= {Tr}(....\rhat)/ {Tr}\rhat .
\end{equation}
The statistical operator $\rhat$ describes the state of the system.
 It can be a pure ensemble or a mixed ensemble depending
on the statistical nature of the system. For a quantum statistical
system, the grand canonical ensemble is usually used
\cite{r5}, whereas  for
heavy-ion collisions $\rhat$ should be specified by the
groundstates of two colliding nuclei \cite{r10,r15,r23}
properly boosted towards each other.
Since in this paper we are essentially
dealing with the formal aspects of the theory,
$\rhat$ is left unspecified furtheron.

The equations of motion for green's functions can be derived directly
from the Heisenberg equations (2.3a,b). They read:
\begin{equation}
\label{e209}
[i\partial_{t_1}-\that(1)]G(1;1')=\delta^4(1,1')-i\int d^4 2 \
\Vhat(1,2)G^{(2)}(1,2;1'2^+),
\end{equation}
\[ [i\partial_{t_1}-\that(1)]G^{(2)}(1,2;1',2')=
\delta^{(4)}(1,1')G(2;2')-\delta^{(4)}(1,2')G(2;1') \]
\begin{equation}
\label{e10}
-i\int d^{4}3  \ \Vhat(1,3) G^{(3)}(1,2,3;1',2',3^+),
\end{equation}
 \[ .......................\]
\[ [i\partial_{t_1}-\that(1)]G^{(n)}(1,...n;1',...n')=
\sum_{j=1}^n\delta^{(4)}(1,j')(-)^{j-1}G^{(n-1)}(2,...n;1'...
(j-1)',(j+1)',...n')\]
\begin{equation}
\label{e211}
 -i\int d^{4}(n+1)\Vhat(1,n+1)G^{(n+1)}(1,...(n+1);1',...(n+1)^+).
\end{equation}
where $t_{(n+1)^\pm}=t^\pm_{(n+1)}=t_{(n+1)}\pm\varepsilon$, and
$\varepsilon$ is infinitesimally positive. Similar equations of motion
can be obtained for any 4-d coordinates $(x_i)$ and $(x'_i)$.
Eqs.(2.9-11) are known as the  Martin-Schwinger-Kadanoff
hierarchy (MSK) \cite{r4} which provides
a complete set of equations of motion for the unlinked Green's
functions $G^{(n)}$ and allows a nonperturbative treatment of the
quantum many-body problem if an appropriate truncation scheme is
provided additionally. The MSK equation are integro-differential equations
which have to be supplemented by proper
initial conditions and boundary conditions. In case of equilibrium
systems, the problem has been discussed by Kadanoff and Baym \cite{r5}
in detail.

Two obvious features can be observed from eqs.(2.9-11): (i) They are
coupled linear equations,  (ii) the equation of motion for $G^{(n)}$
is coupled to $G^{(n-1)}$ and $G^{(n+1)}$.  Since all the
many-body green's functions are treated on an equal footing,
 one cannot set any
of them simply to zero. This means that eqs.(2.9-11) themselves do not
provide a natural or transparent truncation scheme.
 Yet, another shortcoming of eqs.(2.9-11) is related to
the large amount of unlinked terms and thus an
unnecessary repetition of information. As we know from diagrammatic
perturbation theory, the n-body green's function
contains a large number of disconnected diagrams, while the useful
information is contained only in the linked irreducible diagrams.

 To get rid of these shortcomings, as in the case of density
matrices \cite{r17}, one has to separate correlation green's functions
$G^{(n)}_c$ explicitly from $G^{(n)}$ and transform the equations of
motion for $G^{(n)}$ to those for $G^{(n)}_c$.

\subsection{\it Correlation Green's Functions and their Equations
of Motion}
There are two major steps to derive  the correlation dynamics: (i) the
separation of many-body correlation green's functions and (ii) the
transformation of equations of motion. The first step can be
accomplished by a generalization of the method used previously in Ref.
\cite{r17}. It results in considering the particle coordinates as
4-dimensional space-time variables, $x_i=(\bx_i,t_i)$ in the respective
cluster expansion, $i.e.$
\begin{equation}
\label{e212}
G^{(2)}(1,2;1',2')=G^{(2)}_c(1,2;1',2')+ \A p_{(1',2')} \S p_{(2; 2')}
G(1;1')G(2;2'),
\end{equation}
\begin{eqnarray}
\label{e213}
\lefteqn{G^{(3)}(1,2,3;1',2',3')=G^{(3)}_{c}(1,2,3;1',2',3')}
\nonumber \\
 &&+\A p_{(1',2',3')}\S p_{(2,3;2',3')}[G(1;1')G^{(2)}(2,3;2',3')+
 G^{(2)}_c(1,2;1'2')G(3;3')],
\end{eqnarray}
 \[ ....................  \]
\begin{eqnarray}
\label{e214}
\lefteqn{ G^{(n)}(1,...n;1',...n')=G^{(n)}_c(1,...n;1',...n') + }
\nonumber \\
 && \A p_{(1'...n')} \S p_{(2,...n;2',...n')} \sum_{k=1}^{n-1}
     G^{(k)}_c(1,..k;1'..k')G^{(n-k)}((k+1),..n;(k+1)',..n'),
\end{eqnarray}
where $G(1;1')=G^{(1)}(1;1')$. \(\S p_{(..;..)} \) denotes symmetrization
of  particle  pairs \((x_i,x'_i)\) and
\((x_j,x'_j)\).  \(\A p_{(...)}\) denotes anti-symmetrization of
 variables $x_i'$ and $x_j'$. In combining the operations
\(\S p\) and \(\A p\), the repeated terms are omitted in case of fermions
as shown in \cite{r17}. The
cluster expansion (2.12-14) is highly nonlinear. It is evident
that \(G^{(n)}\) contains all possible correlations among $n$
particles. The properties of eqs.(2.12-14) are as follows:

1. $G^{(n)}$ can be expanded according to any variable pair
$(x_i,x_i')$. Due to symmetrization and anti-symmetrization
operations, all the expansions are equivalent.

2. $G^{(n)}_c$ and $ \A p \S p  G^{(k)}_c G^{(n-k)}$ possess the same
symmetry as $G^{(n)}$. Yet, $\A p$ and $\S p$ have following
properties,
\begin{equation}
\label{e215}
  \A p_{(...)_1} \S p_{(..;..)_1} \A p_{(...)_2} \S p_{(..;..)_2}
= \A p_{(...)_1} \S p_{(..;..)_1},
\end{equation}
if $(...)_1\supseteq (...)_2$ and $(..;..)_1\supseteq (..;..)_2$.

3. $G^{(n)}$ contains all possible correlations among $n$ particles.

The advantage of eq.(2.14) is that the particle variable $x_1$ is
fixed in $G^{(k)}_c$. This will facilitate the derivation of equations
of motion. It can be proven that eqs.(2.12-14)
 are consistent with the
corresponding relations between $G^{(n)}$ and $G^{(n)}_c$ as obtained
from the generating functional technique;  the correlation green's
functions $G^{(n)}_c$ are simply the connected green's functions.

For reasons of transparency we present the lowest
order cluster expansions,
\begin{equation}
\label{e216}
G^{(2)}(1,2;1',2')= G(1;1')G(2;2') -G(1;2')G(2;1') +
 G^{(2)}_c(1,2;1',2'),
\end{equation}
\begin{eqnarray}
\label{e217}
\lefteqn{G^{(3)}(1,2,3;1',2',3')=G^{(3)}_c(1,2,3;1',2',3')}
\nonumber \\
&&+G(1;1')G(2;2')G(3;3')-G(1,2')G(2;1')G(3;3')-G(1;3')G(2;2')G(3;1')
\nonumber \\
&&-G(1;1')G(2;3')G(3;2')+G(1;2')G(2;3')G(3;1')+G(1;3')G(2;1')G(3;2')
\nonumber \\
&&+G(1;1')G^{(2)}_c(2,3;2',3')-G(1;2')G^{(2)}_c(2,3;1',3')-G(1;3')
G^{(2)}_c(2,3;2',1')
\nonumber \\
&&+G^{(2)}_c(1,2;1',2')G(3;3')-G^{(2)}_c(1,2;3',2')G(3;1')-
G^{(2)}_c(1,2;1',3')G(3;2')
\nonumber \\
&&+G^{(2)}_c(1;3;1',3')G(2;2')-G^{(2)}_c(1,3;2',3')G(2;1')-
G^{(2)}_c(1,3;1',2')G(2;3').
\end{eqnarray}

{}From the above discussion and explicit construction it is obvious that the
correlation Green's functions $G^{(n)}_c$ represent the linked
diagrams, describing the cluster structure of the $n$-particles.
Since all $G^{(n)}$ consist of the products of linked Green's
functions, $G^{(n)}_c$ are the basic elements of our theory which contain
the necessary and sufficient information. Moreover, by introducing
$G^{(n)}_c$, all the unlinked terms in the equations of motion will
be canceled and the equations of motion for $G^{(n)}_c$ will be
simplified substantially.

Having accomplished the cluster expansion of $G^{(n)}_c$, the next step is
to derive equations of motion for $G^{(n)}_c $. This is a complex task
in the general case and  the lengthy mathematical manipulation
 is shifted to Appendix A. However,
for the lowest order equations, the derivation is straightforward.
Inserting eqs.(2.12-13) into eqs.(2.9-11) we obtain the equations of
motion for $G, \ G^{(2)}_c$ and $G^{(3)}_c$,
\begin{eqnarray}
\label{e218}
\lefteqn{ [i\partial_{t_1}-\that(1)]G(1;1')=\delta^{(4)}(1;1') }
\nonumber\\
&&-i\int d2\Vhat(1,2)
[G^{(2)}_c(1,2;1',2^+)+G(1;1')G(2;2^+)-G(1;2)G(2;1')],
\end{eqnarray}
\begin{eqnarray}
\label{e219}
\lefteqn { [i\partial_{t_1}-\that(1)]G^{(2)}_c(1,2;1',2')= }
\nonumber\\
&&-i\int d3 \
\Vhat(1,3)\big [-G^{(2)}(1,3;1',2')G(2;3)+G^{(3)}_c(1,2,3;1',2',3^+)
\nonumber \\
&&+G^{(2)}_c(1,2;1',2')G(3;3^+)-G^{(2)}_c(3,2;1',2')G(1;3)+
G^{(2)}_c(2,3;2',3^+)G(1;1')
\nonumber \\
&&-G^{(2)}_c(2,3;1',3^+)G(1;2')+G^{(2)}_c(1,2;2',3)G(3;1')-
G^{(2)}_c(1,2;1',3)G(3;2')\big ],
\end{eqnarray}
\begin{eqnarray}
\label{e220}
\lefteqn { [i\partial_{t_1}-\that(1)]G^{(3)}_c(1,2,3;1',2',3')=
-i\int d4 \ \Vhat(1,4)\big [G^{(2)}_c(2,3;4,1')G^{(2)}(1,4;3',2') }
 \nonumber\\
&&-G^{(2)}_c(2,3;4,3') G^{(2)}(1,4;1',2')-
G^{(2)}_c(2,3;2',4)G^{(2)}(1,4;1',3')
 \hspace{3cm} \nonumber\\
&&-G^{(2)}_c(1,2;1',2')G(3;4)G(4;3')+
G^{(2)}_c(1,2;1',3')G(3;4)G(4;2') \hspace{3cm} \nonumber\\
&&+G^{(2)}_c(1,2;3',2')G(3;4)G(4;1')
-G^{(2)}_c(1,3;1',3')G(2;4)G(4;2') \hspace{3cm} \nonumber\\
&&+G^{(2)}_c(1,3;1',2')G(2;4)G(4;3')
+G^{(2)}_c(1,3;2',3')G(2;4)G(4;1')
 \hspace{3cm} \nonumber\\
&&+G^{(2)}_c(4,2;1',2')G(3;4)G(1;3')
-G^{(2)}_c(4,2;1',3')G(3;4)G(1;2') \hspace{3cm} \nonumber\\
&&-G^{(2)}_c(4,2;3',2')G(3,4)G(1;1')
+G^{(2)}_c(4,3;1',3')G(2;4)G(1;2') \hspace{3cm} \nonumber\\
&&-G^{(2)}_c(4,3;1',2')G(2;4)G(1;3')
-G^{(2)}_c(4,3;2',3')G(2;4)G(1;1')  \hspace{3cm}
\nonumber\\
&&+G^{(2)}_c(1,2;1',2')G^{(2)}_c(3,4;3',4^+)
-G^{(2)}_c(1,2;1',3')G^{(2)}_c(3,4;2',4^+) \hspace{3cm} \nonumber\\
&&-G^{(2)}_c(1,2;1',4)G^{(2)}_c(3,4;3',2')
+G^{(2)}_c(1,2;3',4)G^{(2)}_c(3,4;1',2') \hspace{3cm} \nonumber\\
&&-G^{(2)}_c(1,2;3',2')G^{(2)}_c(3,4;1',4^+)
-G^{(2)}_c(1,2;4,2')G^{(2)}_c(3,4;3',1')  \hspace{3cm}
\nonumber\\
&&+G^{(2)}_c(1,3;1',3')G^{(2)}_c(2,4;2',4^+)
-G^{(2)}_c(1,3;1',2')G^{(2)}_c(2,4;3',4^+) \hspace{3cm} \nonumber\\
&&-G^{(2)}_c(1,3;1',4)G^{(2)}_c(2,4;2',3')
+G^{(2)}_c(1,3;2',4)G^{(2)}_c(2,4;1',3')  \hspace{3cm} \nonumber\\
&&-G^{(2)}_c(1,3;2',3')G^{(2)}_c(2,4;1',4^+)
-G^{(2)}_c(1,3;4,3')G^{(2)}_c(2,4;2',1')   \hspace{3cm}
\nonumber\\
&&+G^{(3)}_c(1,2,3;1',2',3')G(4;4^+)
-G^{(3)}_c(1,2,3;4,2',3')G(4;1') \hspace{3cm} \nonumber\\
&&-G^{(3)}_c(1,2,3;1',4,3')G(4;2')
-G^{(3)}_c(1,2,3;1',2',4)G(4;3') \hspace{3cm} \nonumber\\
&&+G^{(3)}_c(4,2,3;4^+,2',3')G(1;1')
-G^{(3)}_c(4,2,3;1',2',3')G(1;4) \hspace{3cm}
\nonumber\\
&&-G^{(3)}_c(4,2,3;4^+,1',3')G(1;2')
-G^{(3)}_c(4,2,3;4^+,2',1')G(1;3') \hspace{3cm} \nonumber\\
&&-G^{(3)}_c(1,4,3;1',2',3')G(2;4)
-G^{(3)}_c(1,2,4;1',2',3')G(3;4) \hspace{3cm} \nonumber\\
&&+G^{(4)}_c(1,2,3,4;1',2',3',4^+)\big ].
\end{eqnarray}
To obtain the equations of motion for $G^{(n)}_c$ in general, it is
helpful to examine the structure of the r.h.s.  of eqs.  (2.18-20). The
concepts of linked and unlinked terms introduced in Ref. \cite{r17} are
very useful in this respect.
Though most of the arguments follow closely Ref. \cite{r17} we repeat the
arguments in this paper just for clarity:
A term consisting of a product of several factors is called ${\bf unlinked}$,
if one factor contains variables which do not appear in the others;
otherwise, it is called ${\bf linked}$. In a linked term, each factor contains
at least one variable which appears also in the other factors. In the
language of Feynman  the diagrams of linked terms are
connected topologically while the diagrams of unlinked terms are
disconnected.

In the above terminology, the structure of
eqs.(2.18-20) can be described as follows:

(i) All the terms on the
r.h.s. of eqs.(2.18-20) are linked either due to interactions (called
interaction correlations), or due to Pauli anti-symmetrization
(called Pauli correlations);

(ii) the two-body interactions and Pauli
anti-symmetrization can connect at most three factors; for n-body
correlations, there are only three types of terms like
\mbox{${Tr}_{(n+1)} \Vhat(1,n+1)G^{(n+1)}_c$},
\mbox{${Tr}_{(n+1)}\Vhat(1,n+1)G^{(k)}_cG^{(l)}_c\delta_{k+l,n+1}$},
and  \mbox{${Tr}_{(n+1)}\Vhat(1,n+1)G^{(k)}_cG^{(l)}_cG^{(m)}_c
\delta_{k+l+m,n+1}$};

(iii) the r.h.s. and l.h.s. of eqs.(2.18-20)
possess identical symmetry with respect to permutation of particle
variables;

(iv) in eqs.(2.18-20), $\int dn $ denotes integration over continuous
variables and summation over discontinuous variables of particle $n$
(see the definition of the trace operation,eq.(2.23)).

Following the above observations the equation of motion for $G^{(n)}_c$
can be written as
\begin{eqnarray}
\label{e221}
\lefteqn{[i\partial_{t_1}-\that(1)]G^{(n)}_c(1,...n;1',...n')=}
\nonumber\\
&&-i\int d(n+1)\Vhat(1,n+1)
\big \{G^{(n+1)}_c(1,...n,n+1;1',...n',(n+1)^+)
\nonumber\\
&&+\A p_{(1',...n')}\S p_{(2,..n;2',..n')}\big [\sum_{k=1}^n
G^{(k)}_c(1,..k;1'..k')   \nonumber\\
&&\cdot G^{(n+1-k)}_c((k+1),..n,(n+1);(k+1)',..n',(n+1)^+)
\nonumber\\
&&-\sum_{k=1}^nG^{(k)}_c(1,..k;1',..(k-1)',(n+1)^+)
G^{(n+1-k)}_c((k+1),..n,(n+1);(k+1)',..n',k')
\nonumber\\
&&-\sum_{k=2}^nG^{(k)}_c(1,..(k-1),(n+1);1',...k')
G^{(n+1-k)}_c((k+1),..n,k;(k+1)',..n',(n+1)^+)
\nonumber\\
&&-\sum_{k=1}^{n-1}\sum_{p=1}^{n-k}G^{(k)}_c(1,..k;1',..k')
G^{(p)}_c((n+1),(k+2),..(k+p);(k+1)',(k+2)',..(k+p)')
\nonumber\\
&& \cdot  G^{(n+1-k-p)}_c((k+p+1),..n,(k+1);
          (k+p+1)',..n',(n+1)^+)\big ] \big \},(n\geq 2)
\end{eqnarray}
which is proven rigorously in Appendix A. The above equation can
be rewritten in a more compact form, namely
\[ [i\partial_{t_1}-\that(1)]G^{(n)}_c=
  -i\big [{Tr}_{(n+1)'=(n+1)^+}\Vhat(1,n+1) \]
\begin{equation}
\label{e222}
\cdot  AS_{(n+1)}
 \sum_{k \geq l \geq m=0}G^{(k)}_cG^{(l)}_cG^{(m)}_c
 \delta_{k+l+m,n+1}\big ]_L, (n\geq 2)
\end{equation}
where the notation $AS_{(n+1)}$ is an abbreviation of $\A p\S p$ and
$[... ]_L$ denotes linked terms as in Ref. \cite{r17}.  Here it is
assumed that $G^{(0)}_c=1, G^{(1)}_c=G^{(1)}=G$. The trace operation
is defined as first taking the equal time limit, then integrating over
space-time variables $x$ and summing over internal degrees of freedom
(e.g. spin-isospin variables)
$\alpha$, \ie
\begin{equation}
\label{e223}
 {Tr}_{(n+1)'=(n+1)_+}\equiv
 \sum_{\alpha_{n+1}}\int_{t'_{n+1}=t^+_{n+1}}d^4x_{n+1}.
\end{equation}
Because of the symmetry of $G^{(n)}$ and $G^{(n)}_c$, it is easy to
write down equations of motion for any space-time vaviables $x_j$
and $x'_j$,}
\[ [i\partial_{t_j}-\that(j)]G^{(n)}_c=-i\big[{Tr}_{(n+1)'=(n+1)^+}
\Vhat(j,n+1) \]
\begin{equation}
\label{e224}
\cdot AS_{(n+1)} \sum_{k\geq l\geq m=0}
G^{(k)}_cG^{(l)}_cG^{(m)}_c\delta_{k+l+m,n+1}\big ]_L, (n\geq 2)
\end{equation}
and
 \[ [i\partial_{t'_1}+\that(1')]G(1;1')=
   -\delta^{(4)}(1,1')+ \]
\begin{equation}
\label{e225}
 i\int d2\Vhat(1',2)
     [G(1;1')G(2^-;2)-G(1;2)G(2;1')+G^{(2)}_c(1,2^-;1',2)],
\end{equation}
\begin{eqnarray}
\label{e226}
\lefteqn{[i\partial_{t'_j}+\that(j')]G^{(n)}_c=
  i\big [{Tr}_{(n+1)=(n+1)^{'-}} \Vhat(j',(n+1)')}
\nonumber\\
&& \cdot AS_{(n+1)}\sum_{k\geq l\geq m=0}G^{(k)}_cG^{(l)}_c
G^{(m)}_c\delta_{k+l+m,n+1}\big ]_L, (n\geq 2).
\end{eqnarray}
Eqs.(2.18) and (2.21) constitute a set of basic equations of motion for
many-body correlation green's functions $G^{(n)}_c$ ( not $G^{(n)}$).

In comparing with the MSK hierarchy, the advantages  of the above
equations are as follows:

 (i) The cluster expansion has been
incorporated into the equations of motion systematically up to
all orders and the unlinked terms have been removed;

(ii) the basic physical quantities are
many-body correlation green's functions $G^{(n)}_c$ (not $G^{(n)}$)
which contain sufficient and all
necessary information for the description of a many-body system; there are no
longer unlinked terms in the equations of motion, i.e.  no
 unnecessary repetition of information;

(iii) they are coupled nonlinear, genuine
nonperturbative equations, reflecting dynamical
feedback processes among correlations of different orders;

 (iv) if
the many-body correlations become weaker with increasing order
(this is the case for most physical systems,
especially for nuclear physics), the above hierarchy of
equations provides a natural truncation scheme and each truncation
leads to a nonperturbative approximation;

 (v) as will be  proven below, the
equal time limit of the above equations leads to the familiar
correlation dynamics of density matrices.

\zeqn
\section{The Equal Time Limit }

A generalized theory should include its special theory as a limiting
case. Since the green's function theory is more general than the
density-matrix theory it
should include the correlation dynamics of density matrices \cite{r17}
as a special case. This section is devoted  to this particular proof.

According to the usual notation, the n-body green's function in
the so-called normal equal time limit($NETL$):
\[
 t'_1> t'_2> ....>t'_n> t_n>....>t_2> t_1 \]
leads to the n-body density matrix,
\begin{equation}
\label{e301}
[i^nG^{(n)}(1,...n;1',...n')]_{NETL}= (-)^n\rho_n(1,..n;1'...n';t).
\end{equation}
Correspondingly, the n-body correlation green's function reduces to
the n-body correlation density matrix,
\begin{equation}
\label{e302}
[i^nG^{(n)}_c(1,...n;1',...n')]_{NETL}= (-)^n C_n(1,...n;1',...n';t).
\end{equation}
Rather than the $NETL$, other equal time limits of $G^{(n)}$ will
result in extra terms which consist of unlinked products of
$\delta$-functions and lower order density matrices. As an example,
we write down the one-body green's function within two
different equal time limits, i.e. the normal and the anti-normal equal time
limits,
\aeqn
\begin{equation}
\label{e303}
[iG(1;1')]_{t'_1=t^+_1=t}=-\rho(1;1';t),
\end{equation}
\neqn
\begin{equation}
[iG(t;t')]_{t'_1=t^-_1=t}= \delta^{(3)}(1;1')-\rho(1;1';t).
\end{equation}
\zeqn
\noindent
In Appendix B we explicitely proove the following statement:
the many-body correlation green's functions possess the identical equal
time limit irrespective of the time order taken, i.e.
\begin{eqnarray}
\label{e304}
\lefteqn{[i^nG^{(n)}_c(1,..n;1',...n')]_{\mbox{any equal time limit}}}
\nonumber\\
&& =[i^nG^{(n)}_c(1,...n;1',...n')]_{NETL}=
(-)^nC_n(1,...n;1',...n';t),(n\geq 2).
\end{eqnarray}
We note  that eq.(3.4) is quite different from eq.(3.2)
which is easy to prove and also well known. However,
equation (3.4)  states, that
$i^nG^{(n)}_c$, under ${\bf any}$ equal time limit (not simply the normal
equal time limit as discussed in the literaure), is equal to
$(-1)^nC_n$. This proof is no longer trivial and due to reasons of complexity
shifted to  Appendix B.

The above results can be understood intuitively as follows. Since
different equal time limits only make a difference for unlinked
terms, however, do not affect linked terms, the equal time limit of
$G^{(n)}_c$ is unique  and independent of the time ordering. Graphically
speaking, linked diagrams are connected in 4-dimensional space-time;
the equal time limit just deforms the connected diagrams in a continuous way
but does not change their topology.
We will see later that eq.(3.4) is a key relation for establishing
the relation between the two formalisms in the equal time limit.

Taking the normal equal time limit, from eqs.(2.18) and (2.25) we have
\begin{eqnarray}
\label{e305}
\lefteqn{i\partial_t \rho(1;1';t)= [\that(1)-\that(1')]\rho(1;1';t)
+{tr}_{(2'=2)}[\v(1,2)-\v(1',2')]}\nonumber\\
&&\cdot [\rho(1;1';t)\rho(2;2';t)
-\rho(1;2';t)\rho(2;1';t)+C_2(1,2;1',2';t)];
\end{eqnarray}
where the trace ${tr}_{(n+1)'=(n+1)}$ does not include integration over
time $t$. For $n\geq 2$, under the $NETL$ we have from eqs.(2.24)
and (2.26),
\begin{equation}
\label{e306}
i\partial_t C_n(1,...n;1',...n';t)=\sum_{j=1}^n[I(j)-I'(j')],
\end{equation}
where
\aeqn
\begin{eqnarray}
\label{e307}
\lefteqn{I(j)=\that(j)C_n(1,...n;1',...n';t)
+\big[{tr}_{(n+1)'=(n+1)}\v(j,n+1)
 AS_{(n+1)}}\nonumber\\
 &&\cdot \sum_{k\geq l\geq m=0}
(-i)^{k+l+m}G^{(k)}_cG^{(l)}_cG^{(m)}_c
 \delta_{k+l+m,n+1}\big]_L
\mid_{(t'_1,..t'_n t_n..t_{(j+1)}t'_{(n+1)}t_{(n+1)} t_j..t_1)=t},
\end{eqnarray}
\neqn
and
\begin{eqnarray}
\lefteqn{I'(j')=\that(j') C_n(1,..n;1',..n';t)
+\big [{tr}_{(n+1)'=(n+1)}\v(j',(n+1)')
 AS_{(n+1)}} \nonumber\\
&&\cdot \sum_{k\geq l\geq m=0}(-i)^{k+l+m}G^{(k)}_cG^{(l)}_cG^{(m)}_c
 \delta_{k+l+m,n+1}\big ]_L
\mid_{(t'_1,..t'_jt_{(n+1)}'t_{(n+1)}t_{(j+1)}'..t'_nt_n..t_1)=t},
\end{eqnarray}
\zeqn
where $(t'_1..t'_nt_n..t_{(j+1)}t'_{(n+1)}t_{(n+1)}t_j..t_1)=t$
denotes a special equal time limit with the time order indicated by
the arguments in the parenthesis.

 In eq.(3.7a), by using eq.(3.4),
$(-i)_k G^{(k)}_c(k\geq 2)$ can be replaced by $C_k$. For
$(-i)G^{(1)}_c(p;q)$, there are two cases:

(i) For \ $-iG^{(1)}_c(p;q'\neq (n+1)^+)$ and
$-iG^{(1)}_c(p\leq j,or,p=n+1;q'=(n+1)^+)$, their equal time
limit is normal as the normal nth-particle equal time limit is taken.
Thus these terms can be replaced by $C_1(p;q)=\rho(p;q;t)$ according
to eq.(3.3a ).

(ii) For $-iG^{(1)}_c(n\geq p>j;q'=(n+1)^+)$,
their equal time limit is anti-normal as the normal nth-particle equal
time limit is taken.

  According to eq.(3.3b) we have
\begin{equation}
\label{e308}
-iG^{(1)}_c(n\geq p>j;q'=
(n+1)^+)=-\delta^{(3)}(p;q'=(n+1)) +\rho(p;(n+1);t).
\end{equation}
For eq.(3.7b) the calculation is similar.

{}From the above discussion we observe that, in the calculation of $I(j)$,
only the terms containing $G^{(1)}_c(n\geq p>j;q'=(n+1)^+)$
need special attention. They are
\begin{eqnarray}
\label{e309}
\lefteqn{\big \{{tr}_{(n+1)}\v(j,n+1)\big [-\sum_{i=j+1}^n(-i)^n
G^{(n)}_c(1..(i-1),(n+1),(i+1)..n;1'...n')(-i)G^{(1)}_c(i;n+1)}
\nonumber \\
&& -\sum_{i=j+1}^n \A p_{(1'..n')} \S
p_{(1..(j-1)(j+1)..(i-1)(i+1)..n;1'..(j-1)'(j+1)'..(i-1)'(i+1)'..n')}
\nonumber\\
&&\cdot \sum_{k=1}^{n-1}(-i)^k
G^{(k)}_c(j,1..(k-1);j',1'..(k-1)') \nonumber\\
&& \cdot (-i)^{n-k}
G^{(n-k)}_c(k..(i-1),(n+1),(i+1)..n;k',..(i-1)',i',(i+1)'..n')\nonumber\\
&&(-i)G^{(1)}_c(i;(n+1)^+)\big ]\big \}_{\mbox{equal time limit}}\nonumber\\
&&  =\sum_{i=j+1}^n \v(j,i)C_n(1...n;1'...n';t) , \hspace{6cm} (I)
\nonumber \\
&& +\sum_{i=j+1}^n\v(j,i)\A p_{(1'..n')}\S
p_{(1..(j-1)(j+1)..(i-1)(i+1)..n;1'..(j-1)'(j+1)'..(i-1)'(i+1)'..n')}
\nonumber \\
&&\times \sum_{k=1}^{n-1}C_k(j,1..(k-1);j'..1'..(k-1)';t)
C_{n-k}(k..i..n;k'..i'..n';t) , \hspace{0.5cm} (II)
\nonumber \\
&& -{tr}_{(n+1)}\v(j,n+1) \A p_(...)\S p_(..;..)
\big [\sum_{i=j+1}^n(C_nC_1(i;n+1;t) \hspace{1.5cm} (III)
\nonumber \\
&&+\sum_{k=1}^{n-1}C_kC_{n-k}C_1(i;n+1;t))\big ],\hspace{6.5cm}(IV)
\end{eqnarray}
where the second term in eq.(3.9) can be written as
\begin{equation}
\big [\sum_{i=j+1}^n\v(j,i)AS_{(n)}\sum_{k=1}^{n-1}C_kC_{n-k}\big]_L.
\end{equation}
and $A$ and $S$ are anti-symmetrization and symmetrization operators
as in Ref. \cite{r17}.  The third and fourth terms of eq.(3.9) are of
the same form as the other terms of eq.(3.7a). By using eqs.(3.9-10),
eq.(3.7a) turns to
\aeqn
\begin{eqnarray}
\label{e311}
\lefteqn{I(j)=\big
[\that(j)+\sum_{i=j+1}^n\v(j,i)\big]C_n+\big[\sum_{i=j+1}^n\v
   (j,i)
AS_{(n)}\sum_{k=1}^{n-1}C_kC_{n-k}\big]_L } \nonumber \\
&&+{tr}_{(n+1)}\big[\v(j,n+1)AS_{(n+1)}\sum_{k\geq l\geq m=0}C_kC_lC_m
\delta_{k+l+m,n+1} \big]_L
\end{eqnarray}
Similarly
\neqn
\begin{eqnarray}
\lefteqn{I'(j')=\big[\that(j')+\sum_{i=j+1}^n\v(j',i')\big]C_n+
\big[\sum_{i=j+1}^n\v(j',i')AS_{(n)}\sum_{k=1}^{n-1}C_kC_{n-k}\big]_L}
\nonumber \\
&&+{tr}_{(n+1)}\big[\v(j',n+1)AS_{(n+1)}
\sum_{k\geq l\geq m=0}
C_kC_lC_m\delta_{k+l+m,n+1}\big]_L
\end{eqnarray}
\zeqn
Inserting eqs.(3.11a,b) into eq.(3.6), we obtain the
equation of motion for $C_n$
\begin{eqnarray}
\label{e312}
\lefteqn{i\partial_tC_n=\big[\sum_{j=1}^n(\that(j)-\that(j'))+
\sum_{i>j=1}^{n-1}(\v(j,i)-\v(j',i'))\big]C_n}  \nonumber \\
&&+\big[\sum_{i>j=1}^{n-1}(\v(j,i)-\v(j',i'))AS_{(n)}
\sum_{k=1}^{n-1}C_kC_{n-k}\big]_L \nonumber \\
&&+{tr}_{(n+1)}\big[\sum_{j=1}^n(\v(j,n+1)-\v(j',n+1))AS_{(n+1)}
\sum_{k\geq l \geq m=0}C_kC_lC_m\delta_{k+l+m,n+1}\big]_L
\end{eqnarray}
Eqs.(3.5) and (3.12) are exactly the equations of motion for many-body
correlation density matrices given in Ref. \cite{r17}.
Thus we have proven that
the correlation dynamics of density matrices is the equal time limit of
our correlation  green's function theory. Since in the green's function
formalism
space and time variables are treated on an equal footing,
 the present formalism can be readily extended
 towards a relativistic correlation dynamics of
quantum fields.  It should be pointed out explicitly that though eq.(3.12)
 is given already in Ref. \cite{r17},
 the relationship between eqs.(2.24, 2.26) and eq.(3.12) here is established
for the first time.

Before concluding this section, we like to make a brief remark
on conservation laws. Since  the average of physical observables can
be calculated from  the many-body green's functions in the normal
equal time limit \cite{r2,r4}, our conclusions concerning conservation
laws in Ref. \cite{r17} are also applicable to the present formalism: i.e.,
the conservation of one-body quantities is independent on the
approximation for $G^{(2)}_c$, while the conservation of
two-body quantities is irrespective on approximations for
$G^{(3)}_c$. For a nuclear system, the most important conserved
quantities are the total Fermion number,
the total linear momentum, the total angular momentum (which are
one-body operators), and the total energy (which in nuclear physics
is a two-body operator). Thus the two-body correlation
dynamical equations preserve all above quantities.

\zeqn
\section{Lowest Order Truncations and Trace Relations}

Since the limiting cases have been extensively
discussed in the literature \cite{r17} - \cite{r21}, \cite{r26,r27,r30,r31},
we only quote the relation to conventional many-body theories for completeness
without presenting the actual proofs again.

As in  case of the density-matrix
formalism, the lowest order truncations of the correlation dynamics
of green's functions lead to mean-field theory (Dyson equation and
TDHF). A truncation on the  two-body level leads to the Bethe-Salpeter equation
(summation of ladder diagrams for short-range correlations) and to the Dyson
equation for the polarization operator (summation of ring diagrams for
long-range correlations) as two limiting cases.
A general  theory including nonperturbatively short-range and
long-range correlations as well as their mutual dynamical interactions is
provided in the limit $C_3 = 0 $ . This truncation scheme includes in
a dynamically way all lowest order parquet diagrams in the stationary
limit.  Furthermore, as discussed in Ref. \cite{r17},
a lowest order truncation on the three-body level leads to
Faddeev-like equations in the form of green's functions.
Thus our present formulation is a transparent, compact and time-dependent
generalization of familiar many-body approaches which are regained within
well defined limits. We note that a green's
function correspondence of the time-dependent G-matrix theory (TDGMT)
within correlation dynamics \cite{r18} has been obtained also in
the present formalism \cite{r33}.

\subsection{Restoration of trace relations}

In the equal time limit the fundamental trace relations for a fermion system
of particle number A, $i.e.$
\begin{equation}
\rho _n = \frac{1}{(A-n)!} \  tr_{(n+1, \dots, A)} \rho _A =
\frac{1}{(A-n)} \  tr_{(n+1)} \rho _{n+1} , \label{e4.1}
\end{equation}
have to be fulfilled which impose severe constraints on $\rho_n$ or
equivalently on
$G^{(n)}_{NETL}$ according to eq. (3.1). Inserting the cluster expansion
(2.12) - (2.14) we obtain the secondary trace relations (n $\leq$ 4)
\begin{equation}
tr_{(2=2')} C_2(12,1'2') = - tr_{(2=2')}
(1/A-P_{12})\rho (22')\rho (11') , \label{e4.2}
\end{equation}
\begin{equation}
tr_{(3=3')} C_3(123,1'2'3') =
 -tr_{(3=3')} (2/A -P_{13}-P_{23}-P_{1'3'}-P_{2'3'})\rho (33')C_2(12,1'2'),
\label{e4.3}
\end{equation}

\begin{eqnarray}
\label{e4.5}
\lefteqn{tr_{(4=4')} C_4(1234,1'2'3'4') = }   \nonumber \\
&& -tr_{(4=4')} (3/A -P_{14}-P_{24}-P_{34}-P_{1'4'}-P_{2'4'}-P_{3'4'})
\nonumber \\
&& \times \rho (44') C_3(123,1'2'3') \nonumber \\
&& +tr_{(4=4')} (P_{14}+P_{24}+P_{1'4'}+P_{2'4'})
C_2(34,3'4') C_2(12,1'2') \nonumber \\
&& +tr_{(4=4')} (P_{34}+P_{3'4'}) C_2(24,2'4') C_2(13,1'3') \nonumber \\
&& -tr_{(4=4')} (C_2(12,3'4') C_2(34,1'2') + C_2(23,1'4') C_2(14,2'3')
\nonumber \\
&& + C_2(24,1'3') C_2(13,2'4'))
\end{eqnarray}
for  correlation functions $C_n$ or equivalently for the correlated
green's functions $G^{(n)}_c$ according to (3.2). As first noted in ref.
\cite{r30} these secondary trace relations
may be violated in time when adopting  $C_3 = 0$
or $C_4 = 0$, respectively. In the above equations $P_{ij}$ denotes the
exchange operator between particle $i$ and $j$ and acts on all terms to
the right.

In order to conserve the trace relations (\ref{e4.1}) or (4.2) - (4.4)
at all times up to a specific order  one has to
introduce explicit three-body correlations $C_3$ in the two-body equation
or four-body correlations $C_4$ in the three-body equation
by a proper functional of $\rho $, $C_2$ or $C_3$ \cite{r31}.
Corresponding functionals are generated by exchanges with an additional
particle of the A-body system, $i.e.$ (dropping the explicit time
dependence)
\begin{eqnarray}
\label{e4.6}
\lefteqn{C_3(123,1'2'3') = }  \nonumber \\
&& - \frac{1}{(A-1)} tr_{(4=4')}
(2/A-P_{14}-P_{24}-P_{1'4'}-P_{2'4'}) \rho _2(34,3'4') C_2(12,1'2') ,
\end{eqnarray}

\begin{eqnarray}
\label{e4.7}
\lefteqn{C_4(1234,1'2'3'4') = }  \nonumber \\
&& - \frac{1}{(A-1)} tr_{(5=5')}
(3/A-P_{15}-P_{25}-P_{35}-P_{1'5'}-P_{2'5'}-P_{3'5'}) \nonumber \\
&&    \times \rho_2(45,4'5') C_3(123,1'2'3') \nonumber \\
&& - \frac{1}{2} tr_{(5=5')} (P_{15}+P_{25}+P_{1'5'}+P_{2'5'})
\nonumber \\
&& \times \{C_3(345,3'4'5') - (P_{34}+P_{54}+P_{3'4'}+P_{5'4'}) \rho(44')
C_2(35,3'5') \} C_2(12,1'2') \nonumber \\
&& - \frac{1}{2} tr_{(5=5')} (P_{35}+P_{3'5'}) \nonumber \\
&& \times \{C_3(245,2'4'5') - (P_{24}+P_{54}+P_{2'4'}+P_{5'4'}) \rho(44')
C_2(25,2'5') \} C_2(13,1'3') \nonumber \\
&& + \frac{1}{2} tr_{(5=5')} P_{2'5'} \nonumber \\
&& \times \{C_3(345,1'4'5') - (P_{34}+P_{54}+P_{1'4'}+P_{5'4'}) \rho(44')
C_2(35,1'5') \} C_2(12,3'2') \nonumber \\
&& + \frac{1}{2} tr_{(5=5')} P_{3'5'} \nonumber \\
&& \times \{C_3(145,2'4'5') - (P_{14}+P_{54}+P_{2'4'}+P_{5'4'}) \rho(44')
C_2(15,2'5') \} C_2(23,1'2') \nonumber \\
&& + \frac{1}{2} tr_{(5=5')} P_{3'5'} \nonumber \\
&& \times \{C_3(245,1'4'5') - (P_{24}+P_{54}+P_{1'4'}+P_{5'4'}) \rho(44')
C_2(25,1'5') \} C_2(23,1'2').
\end{eqnarray}
\noindent
By taking trace of (\ref{e4.6}) with respect to particle 3 we regain
(\ref{e4.3})
and by taking trace of (\ref{e4.7}) with respect to particle 4 we regain
(\ref{e4.5}) when using (\ref{e4.3}).

It must be noted that the explicit
expressions (\ref{e4.6}) and (\ref{e4.7}) are not fully
antisymmetric with respect to all particle exchanges, e.g. the expression
(\ref{e4.6}) is antisymmetric with respect to exchanges (1,2) and (1',2'),
however, not with respect to (1,3) and (2,3). Furthermore, the trace relations
so far are only fulfilled for particle 3 or 4, respectively, but not for
particle 1 or 2. The  expressions for trace relations with respect to
particle $i$ are obtained from (\ref{e4.6}) by operating with $P_{i3}P_{i'3'}$
and from (\ref{e4.7}) by the operation $P_{i4} P_{i'4'}$.

The above considerations now allow to formulate trace conserving theories on
the two- and three-body level also in case of correlated green's functions.
When restricting to the two-body level one has to insert
\begin{equation}
G^{(3)}_c(1,2,3;1',2',3^+) = -i P_{13} P_{1'3'}
\ C_3(123,1'2'3')|_{3'=3^+}  \label{e4.8}
\end{equation}
in eq. (2.19) with $C_3$ from (\ref{e4.6}) whereas in case of
a conserving theory on the three-body level one has to insert
\begin{equation}
G_c^{(4)}(1,2,3,4;1',2',3',4^+) = P_{14} P_{1'4'}
\ C_4(1234,1'2'3'4')|_{4'=4^+}    \label{e4.9}
\end{equation}
in eq. (2.20) with $C_4$ from (\ref{e4.7}).

Due to the lack of a full antisymmetry with respect to all particle
exchanges the respective equations of motion should be averaged over spin
and be applied to spin symmetric systems as in \cite{r31,r34}. As shown
explicitely in \cite{r31,r34} for many-particle systems on the two-body
level the secondary trace relation (\ref{e4.2}) holds very well.
Furthermore, it could be shown that the dynamical influence of the terms
with $C_3$ in the two-body equation is negligible for the dynamical response
of nuclei up to a temperature of a few MeV.

\subsection{Numerical investigations}

The latter findings can be most easily demonstrated by evaluating the
total traces of $C_2, C_3$ and $C_4$ for configurations close to the
groundstate. Since the total traces are basis invariant we choose a
single-particle basis $|\alpha>$ which diagonales the one-body density
matrix, $i.e. \ \rho_{\alpha \alpha'} = n_{\alpha} \delta_{\alpha \alpha'}$.
Within this basis the respective traces lead to (using (\ref{e4.2}) -
(\ref{e4.5}))
\begin{equation}
tr \ C_2 = - \sum_{\alpha} n_{\alpha} (1 - n_{\alpha}), \label{e4.10}
\end{equation}
\begin{equation}
tr \ C_3 = 2 \sum_{\alpha} n_{\alpha} (1 - n_{\alpha})
(1 - 2 n_{\alpha}), \label{e4.11}
\end{equation}
\begin{eqnarray}
\label{e4.12}
\lefteqn{tr \ C_4 = -6 \sum_{\alpha} n_{\alpha} (1 - n_{\alpha})
(1 - 2 n_{\alpha}) + 12 \sum_{\alpha} n_{\alpha}^2 (1 - n_{\alpha})^2
+ 12 \sum_{\alpha \beta} n_{\alpha} n_{\beta} C_{\alpha \beta
\alpha \beta} }  \nonumber \\
&& + \sum_{\alpha \beta \gamma \lambda} \{C_{\alpha \beta \alpha \lambda}
C_{\gamma \lambda \gamma \beta} +
C_{\alpha \beta  \lambda \beta} C_{\gamma \lambda \gamma \alpha}
 + C_{\gamma \beta \gamma \lambda} C_{\alpha \lambda \alpha \beta}
+ C_{\lambda \beta \alpha \beta} C_{\gamma \alpha \gamma \lambda} \nonumber \\
&& + C_{\gamma \beta \lambda \beta} C_{\alpha \lambda \alpha \gamma}
+ C_{\lambda \beta \gamma \beta} C_{\alpha \gamma \alpha \lambda}
- C_{\alpha \beta \gamma \lambda} C_{\gamma \lambda \alpha \beta}
- C_{\gamma \beta \alpha \lambda} C_{\alpha \lambda \gamma \beta}
- C_{\beta \lambda \alpha \gamma} C_{\alpha \gamma \beta \lambda}\}.
\end{eqnarray}
Thus $tr \ C_2$ expresses the fluctuation in the particle number (except for
a minus sign) whereas $tr \ C_3$ describes a $3^{rd}$ order fluctuation. Since
both quantities only depend on $n_{\alpha}$ we can compute them for any given
configuration.

As a first example we investigate a Fermi gas
in equilibrium at finite temperature T with
$n_{\alpha}$ given by the respective Fermi distribution for a system of 40
particles. The numerical results according to (\ref{e4.10}) and (\ref{e4.11})
are displayed in Fig. 1 and have to be compared with $tr \rho_2 = 1560$ and
$tr \rho_3 = 59.280$, respectively. It becomes obvious that the relative
weight of $C_3$ with respect to $\rho_3$ is only very small especially at
low temperature.

In order to obtain an estimate for the values of $tr \ C_4$ according to
(\ref{e4.12}) we, furthermore, need the explicit two-body correlation
matrix elements $C_{\alpha \beta \alpha'\beta'}$ in a given basis. In this
respect we have performed nonperturbative calculations on the two-body
level for $^{40}Ca$ at finite temperature
within the framework of TDDM (cf. \cite{r28,r29,r34,r35}) where the
equation of motion for the two-body correlation matrix is solved numerically
within a time-dependent Hartree-Fock basis. For details on the basis states
and the interaction adopted we refer the reader to Ref. \cite{r29}.

For the particular case shown in Fig. 2 we have initialized the system by
a Hartree-Fock configuration at finite temperature T = 2 MeV and propagated
the system in time for about $10^{-21}$ evaluating at each timestep $tr \ C_2,
tr \ C_3$ and $tr \ C_4$ according to (\ref{e4.10}) - (\ref{e4.12}). The actual
time evolution of $tr \ C_2$ (dashed line), $tr \ C_3$ (solid line $\times 10$)
and $tr \ C_4$ (dotted line) in Fig. 2 is displayed for $t \geq 40 \times
10^{-23} s$ and shows oscillations in time around some asymptotic mean
values. The actual mean numbers for T = 2 MeV
amount to $tr \ C_2 \approx -2.1$ in line with
the Fermi-gas value in Fig. 1, $ tr \ C_3 \approx 0.45$ which is considerably
higher than the corresponding value in Fig. 1, however, still
small compared to A(A-1)(A-2) with A = 40. The four-body trace on average
is $tr \ C_4 \approx 4$ and by orders of magnitude smaller than
A(A-1)(A-2)(A-3).
Comparing the relative weights $tr C_n/tr \rho_n$ we obtain
$1.3 \times 10^{-3}$ for n = 2, $7.6 \times 10^{-6}$ for n = 3 and $1.8 \times
10^{-6}$ for n = 4 indicating a convergence of the cluster expansion. We note
that similar ratios are obtained also at temperatures of 3, 4, and 5 MeV.

\section{Discussion and Outlook}
So far we have generalized the correlation dynamics of density matrices
and established a complete set of dynamical equations for many-body
correlation green's functions. The resulting formalism is
nonperturbative and nonlinear.  It provides a natural truncation
scheme with respect to the order of many-body correlations. In the
lowest order truncations, the main results of the conventional green's
function theory have been obtained.  Besides, the two-body correlation
dynamics contains ladder diagrams and ring diagrams in a compact way.
In general, the correlation dynamics of green's functions provides a
unified and systematic method to treat the quantum many-body problem in a
nonperturbative manner.

Whereas the conventional green's function approach results in a perturbative
calculation of green's functions, our approach rigorously
establishes a set of dynamical equations for correlation green's functions.
The advantage of the present approach resides in that it provides a
complete set of nonperturbative equations and that its truncation
schemes are natural and systematic. Whereas particle number, momentum and
energy are strictly conserved within our formulation the fundamental
trace relations (\ref{e4.1}) are partly violated when truncating on a
specific level. Thus we have explicitely constructed functionals for
the higher order correlated green's functions $G^{(3)}_c$ and $G^{(4)}_c$
which restore the trace relations dynamically for spin-symmetric systems,
$i.e.$ when avering over the spin of the fermions. Numerical studies clearly
demonstrate, furthermore, the convergence of the adopted cluster expansion
at least for nuclear configurations close to the groundstate.

In comparison to the MSK hierarchy our approach
no longer contains unlinked interaction terms which leads to more compact
equations of motion  for the linked Green's functions and provides
 an explicit
treatment of the correlation Green's functions $G^{(n)}_c$, which are
not specified explicitly in the conventional MSK formulation.

The present approach is still in a multi-time formulation. It reduces to
the familiar correlation dynamics of density matrices in the equal time limit.
Though the difference might not be very important in the non-relativistic case,
multi-time green's functions are essential
for a proper description of retardation effects and causality in relativistic
many-body theories. Thus
the present correlation dynamics in a multi-time form is a
necessary step towards a fully relativistic theory of correlation
dynamics, which might tackle the low-energy field theoretical
problem in a rigorous way.

\newpage

\appendix
\noindent\mbox{\bf\Large APPENDIX}
\section{Proof of Equation (2.21)}
We  prove equation (2.21) by induction.
 Since for $n=2$  and  $3$ it reduces to eqs.(2.19) and (2.20),
respectively, equation (2.21) is correct for $2\leq n \leq 3 $.
 Next we  prove that if eq.(2.21) is valid for $2\leq n \leq m $
, then it is valid for $ 2 \leq n \leq m+1 $ . According to
the cluster expansion (2.14) we have
\begin{eqnarray}
\label{a01}
\lefteqn{G^{(m+1)}_c(1...(m+1);1'...(m+1)')=
G^{(m+1)}(1...(m+1);1'...(m+1)')} \nonumber \\
&& -\A p_{(1'..(m+1)')}\S p_{(2..(m+1);2'..(m+1)')}
\sum_{k=1}^m G^{(k)}_c (1..k;1'..k') \nonumber \\
&&\cdot G^{(m+1-k)}((k+1)..(m+1);(k+1)'..(m+1)').
\end{eqnarray}
Operating with $[i\partial_{t_1}-\that(1)]$ from  left on eq.(A.1),
we obtain
\begin{eqnarray}
\label{a02}
\lefteqn{[i\partial_{t_1}-\that(1)]G^{(m+1)}_c(1..(m+1);1'..(m+1)')= }
 \nonumber\\
&& [i\partial_{t_1}-\that(1)]G^{(m+1)}(1..(m+1);1'..(m+1)')
-\A p_{(1'..(m+1)')}\S p_{(2..(m+1);2'..(m+1)')\cdot \nonumber\\
&& \sum_{k=1}^m
 \big[(i\partial_{t_1}-\that(1))G^{(k)}_c(1..k;1'..k')\big]
 G^{(m+1-k)}((k+1)..(m+1);(k+1)'..(m+1)').
\end{eqnarray}
Inserting eq.(2.11) into eq.(A.2),\ using eq.(2.9) for $G^{(1)}=G$ and
eq.(2.21) for $2\leq k\leq m $ and noticing eq.(2.15), we have
\begin{eqnarray}
\label{a03}
\lefteqn{[i\partial_{t_1}-\that(1)]G^{(m+1)}_c(1..(m+1);1'..(m+1)')=
-i\int d(m+2)\Vhat(1,m+2)}\nonumber \\
&& \Big
\{G^{(m+2)}(1..(m+1),(m+2);1'..(m+1)',(m+2)^+)\hspace{5.3cm}(I)\nonumber\\
&&-\A p_{(1'..(m+1)')}\S p_{(2..(m+1);2'..(m+1)')}\Big(\sum_{k=1}^m
\big[G^{(k+1)}_c(1..k,(m+2);1'..k',(m+2)^+)\hspace{1cm}(II)  \nonumber\\
&&+\sum_{p=1}^kG^{(p)}_c(1..p;1'..p')G^{(k+1-p)}_c
((p+1)..k,(m+2);(p+1)'..k',(m+2)^+) \hspace{1.7cm}(III) \nonumber\\
&&-\sum_{p=1}^kG^{(p)}_c(1..p;1'..(p-1)',(m+2)^+)
G^{(k+1-p)}_c((p+1)..k,(m+2);(p+1)'..k',p')\big] (IV) \nonumber \\
&&\cdot G^{(m+1-k)}((k+1)..(m+1);(k+1)'..(m+1)') \nonumber \\
&&-\sum_{k=2}^m\big[\sum_{p=2}^kG^{(p)}_c(1..(p-1),(m+2);1'..(p-1)',p')
\nonumber\\
&&\cdot G^{(k+1-p)}_c((p+1)..k,p;(p+1)'..k',(m+2)^+) \hspace{6.4cm}(V)
\nonumber
\\
&&+\sum_{p=1}^{k-1}\sum_{l=1}^{k-p}G^{(p)}_c(1..p;1'..p')
G^{(l)}_c((p+1)..(p+l-1),(m+2);(p+1)'..(p+l)')\hspace{1.0cm}(VI) \nonumber \\
&&\cdot G^{(k+1-p-l)}_c((p+l+1)..k,(p+l);(p+l+1)'..k',(m+2)^+)\big]
\nonumber \\
&& \cdot G^{(m+1-k)}((k+1)..(m+1);(k+1)'..(m+1)')\Big) \Big \}
\end{eqnarray}
The following multiple summation relations are needed for the further
calculation,
\aeqn
\begin{equation}
\label{a04}
\sum_{k=1}^m\sum_{p=1}^k =\sum_{p=1}^m\sum_{k=p}^m,
\end{equation}
\neqn
\begin{equation}
\sum_{k=2}^m \sum_{p=2}^k =\sum_{p=2}^m\sum_{k=p}^m,
\end{equation}
\neqn
\begin{equation}
\sum_{k=2}^m\sum_{p=1}^{k-1}\sum_{l=1}^{k-p}=
\sum_{p=1}^{m-1}\sum_{l=1}^{m-p}\sum_{k=p+l}^m.
\end{equation}
After having changed the order of summations, eq.(A.3) is simplified
and the third to sixth terms of the r.h.s. are as follows:
\aeqn
\begin{eqnarray}
\label{a05}
\lefteqn{(VI)=-i\int d(m+2) \Vhat(1,m+2)}\nonumber\\
&&\cdot \A p_{(1'..(m+1)')}\S p_{(2..(m+1);2'..(m+1)')}
\sum_{p=1}^m
\sum_{l=1}^{m+1-p}
 [G^{(m+2-p-l)}-G^{(m+2-p-l)}_c ] \nonumber\\
&& (p+l+1..m+1,p+l;(p+l+1)'..(m+1)',(m+2)^+),
\end{eqnarray}
\neqn
\begin{eqnarray}
\lefteqn{(III)+(VI)=i\int d(m+2)\Vhat(1,m+2)\A p_{(1'..(m+1)')}\S
p_{(2..(m+1);2'..(m+1)')}}\nonumber\\
&&\cdot \sum_{p=1}^m\sum_{l=1}^{m+1-p}
G^{(p)}_c(1..p;1'..p')G^{(l)}_c((p+1)..(p+l-1),(m+2);(p+1)'..(p+l)')
\nonumber \\
&&\cdot  G^{(m+2-p-1)}_c((p+l+1)..(m+1),(p+l);(p+l+1)'..(m+2)^+)
\hspace{1cm} (I) \nonumber \\
&&+i\int d(m+2)\Vhat(1,m+2)\A p_{(1'..(m+1)')}\S
p_{(2..(m+1);2'..(m+1)')}\sum_{p=1}^{m+1}G^{(p)}_c(1..p;1'..p')
\nonumber \\
&&\cdot [G^{(m+2-p)}-G^{(m+2-p)}_c]((p+1)..(m+2);(p+1)'..(m+2)^+)
\hspace{1.5cm} (II)
\end{eqnarray}
\neqn
\begin{eqnarray}
\lefteqn{(IV) =-i\int d(m+2)\Vhat(1,m+2)\A p_{(1'..(m+1)')}
\S p_{(2..(m+1);2'..(m+1)')}}\nonumber\\
&&\cdot \sum_{p=1}^{m+1}
 G^{(p)}_c(1..p;1'..(p-1)',(m+2)^+) \nonumber \\
&&\cdot [G^{(m+2-p)}-G^{(m+2-p)}_c]((p+1)..(m+2);
(p+1)'..(m+1)',p'),
\end{eqnarray}
\neqn
\begin{eqnarray}
\lefteqn{(V)=-i\int d(m+2)\Vhat(1,m+2)\A p_{(1'..(m+1)')}
\S p_{(2..(m+1);2'..(m+1)')}}\nonumber\\
&&\cdot \sum_{p=2}^{m+1}
 G^{(p)}_c(1..(p-1),(m+2);1'..p')
 [G^{(m+2-p)}-G^{(m+2-p)}_c]\nonumber\\
 && ((p+1)..(m+1),p;
(p+1)'..(m+1)',(m+2)^+).
\end{eqnarray}
\zeqn
Now we proceed to sum up all the unlinked terms, i.e., the second
term of eq.(A.3), the second term of eq.(A.5b), and the first term of
eq.(A.5c,d). The sum of the four terms amounts to
\begin{equation}
\label{a06}
i\int d(m+2) \Vhat(1,m+2)[G^{(m+2)}-G^{(m+2)}_c](1..(m+1),(m+2);
1'..(m+1)',(m+2)^+),
\end{equation}
which exactly cancels the unlinked terms contained in $G^{(m+2)}$ in
the first term of eq.(A.3). Thus only the linked terms are left in the
r.h.s. of eq.(A.3). We finally obtain

\begin{eqnarray}
\label{a07}
\lefteqn{[i\partial_{t_1}-\that(1)]G^{(m+1)}_c(1..(m+1);1'..(m+1)')=
-i\int d(m+2) \Vhat(1,m+2)  } \nonumber \\
&&\cdot \Big\{G^{(m+2)}_c(1..(m+1),(m+2);1'..(m+1)',(m+2)^+)
+\A p_{(1'..(m+1)')}\S p_{(2..(m+1);2'..(m+1)')}       \nonumber\\
&&\cdot \big[\sum_{p=1}^{m+1}
G^{(p)}_c(1..p;1'..p')G^{(m+2-p)}_c((p+1)..(m+2);(p+1)'..(m+2)^+)
\nonumber \\
&&-\sum_{p=1}^{m+1}G^{(p)}_c(1..p;1'..(p-1)',(m+2)^+)
G^{(m+2-p)}_c((p+1)..(m+2);(p+1)'..(m+1)',p') \nonumber \\
&&-\sum_{p=2}^{m+1}G^{(p)}_c(1..(p-1),(m+2);1'..p')
G^{(m+2-p)}_c((p+1)..(m+1),p;(p+1)'..(m+2)^+) \nonumber \\
&&-\sum_{p=1}^m\sum_{l=1}^{m=1-p}G^{(p)}_c(1..p;1'..p')
G^{(l)}_c((p+1)..(p+l-1),(m+2);(p+1)'..(p+l)') \nonumber \\
&& \cdot G^{(m+2-p-l)}_c((p+l+1)..(m+1),(p+l);(p+l+1)'
..(m+1)',(m+2)^+)\big] \Big \}.
\end{eqnarray}
This is exactly  equation (2.21)  for $n=m+1$ and means that it is
valid for $2 \leq n \leq m+1$.
 From the above calculation we have  proven eq.(2.21) by induction.
The key point of the above proof is to confirm a complete cancellation
of unlinked terms.

\newpage

\section{Proof of Equation (3.4)}
Equation (3.4) is  proven also by induction. Let
$\That(\alpha)(\alpha=(pq),(p'q'),(pq'))$ be a time-ordering operator
such that
\aeqn
\begin{equation}
\label{b01}
\That(p'q')<..\psi^{\dagger}(p')..\psi^{\dagger}(q')..>=
-<..\psi^{\dagger}(q')..\psi^{\dagger}(p')..>,
\end{equation}
\neqn
\begin{equation}
\That(pq)<..\psi(p)..\psi(q)..>=-<..\psi(q)..\psi(p)..>,
\end{equation}
\neqn
\begin{equation}
\That(pq')<..\psi^{\dagger} (q')..;..\psi(p)..>=
 -<..\psi(p)..;..\psi^{\dagger}(q')..>.
\end{equation}
\zeqn
In what follows we shall refer $p$ and $q$  as $\psi(p)$ and $\psi(q)$,
and $p'$ and $q'$ as $\psi^{\dagger}(p')$ and $\psi^{\dagger}(q')$.
We introduce the normal time ordering green's function $G^{(m)N}$
which is defined as the average of the normal product of field operators,
\begin{equation}
\label{b02}
(-i)^mG^{(m)N}(1..m;1'..m')=<\psi^{\dagger}(1')..\psi^{\dagger}
(m')\psi(m)..\psi(1)>.
\end{equation}
Any time ordering green's function $G^{(m)any} $ \ can be obtained by
operating $\That(\alpha)\That(\beta)..\That(\gamma)$ on $G^{(m)N}$,
\begin{equation}
\label{b03}
G^{(m)any}(1..m;1'..m')=\That(\alpha)\That(\beta)..\That
(\gamma)G^{(m)N}(1..m;1'..m').
\end{equation}
Therefore one needs only to study the effect of $\That(\alpha)$.
  Since
\begin{equation}
\label{b04}
\That(pq)G^{(m)N}=\That(p'q')G^{(m)N}=G^{(m)N},
\end{equation}
one concludes that
\begin{equation}
\label{b05}
\That(pq)G^{(m)N}_c=\That(p'q')G^{(m)N}_c=G^{(m)N}_c.
\end{equation}
Thus only $\That(pq')$ needs to be investigated.

 First we examine the effect of $\That(pq')$ on $G^{(1)N}=G^{N}$
and $G^{(2)N}_c$. Let us use $ETL$(Equal Time Limit), $NETL$(Normal
Equal Time Limit), $ANETL$(Anti-Equal Time Limit), and $any-ETL$(any
Equal Time Limit) to denote different equal time limits. For $n=1$,
from eqs.(3.3a,b), we have
\begin{equation}
\label{b06}
\That(pq')G^{N}(p;q')_{ETL}=G(p;q')_{ANETL}=G(p;q')_{NETL}-i\delta^{(3)}
(p,q').
\end{equation}
For $n=2$, we calculate
\label{b07}
\begin{eqnarray}
\lefteqn{\That(pp')(-i)^2G^{(2)N}(pq;p'q')_{ETL}=
\That(pp')<\psi^{\dagger}(p')\psi^{\dagger}(q')\psi(q)\psi(p)>_{ETL}}
\nonumber \\
&&=-<\psi(p)\psi^{\dagger}(q')\psi(q)\psi^{\dagger}(p')>_{ETL}
\nonumber \\
&&=(-i)^2G^{(2)}(pq;p'q')_{NETL}-\delta^{(3)}(p,q')\delta^{(3)}(q,p')-
i\delta^{(3)}(p,q')G(q;p')_{NETL} \nonumber \\
&&-i\delta^{(3)}(q,p')G(p;q')_{NETL}+i\delta^{(3)}(p,p')G(q;q')_{NETL}.
\end{eqnarray}
{}From eq.(2.14) we have, for $n=2$,
\begin{equation}
\label{b08}
G^{(2)N}(pq;p'q')=G^{(2)N}_c(pq;p'q')+G^N(p;p')G^N(q;q')-
G^N(p;q')G^N(q;p'),
\end{equation}
and
\[ G^{(2)}(pq;p'q')_{NETL}=G^{(2)}_c(pq;p'q')_{NETL}+ \]
\begin{equation}
\label{b09}
 G(p;p')_{NETL}G(q;q')_{NETL}-
 G(P;q')_{NETL}G(q;p')_{NETL}.
\end{equation}
Applying $\That(pp')$ on eq.(B.8), we have
\begin{eqnarray}
\label{b10}
\lefteqn{\That(pp')G^{(2)N}(pq;p'q')_{ETL}=
\That(pp')G^{(2)N}_c(pq;p'q')_{ETL}}\nonumber\\
&&+\That(pp')G^N(p;p')_{ETL}G^N(q;q')_{ETL} \nonumber\\
&&-\That(pp')
G^N(p;q')_{ETL}\That(pp')G^N(q;p')_{ETL}.
\end{eqnarray}
Noticing that $\That(pp')G^N(p;p')_{ETL},\That(pp')G^N(p;q')_{ETL}$
and $\That(pp')G^N(q;p')_{ETL}$ are in anti-equal time limit, from
eq.(B.6) we thus have
\begin{eqnarray}
\label{b11}
\lefteqn{\That(pp')G^{(2)N}(pq;p'q')_{ETL}=\That(pp')
G^{(2)N}_c(pq;p'q')_{ETL}} \nonumber \\
&&+[G^{(2)}(pq;p'q')_{NETL}-G^{(2)}_c(pq;p'q')_{NETL}]
-i\delta^{(3)}(p,p')G(q;q')_{NETL} \nonumber \\
&&+i\delta^{(3)}(p,q')G(q;p')_{NETL}
+i\delta^{(3)}(q,p')G(p;q')_{NETL}\nonumber\\
&&+\delta^{(3)}(p,q')\delta^{(3)}(q,p').
\end{eqnarray}
Comparing eqs.(B.7) and (B.11) we obtain
\aeqn
\begin{equation}
\label{b12}
\That(pp')G^{(2)N}_c(pq;p'q')_{ETL}=G^{(2)}_c(pq;p'q')_{NETL}.
\end{equation}
\neqn
Similarly one has
\begin{eqnarray}
\lefteqn{\That(pq')G^{(2)N}_c(pq;p'q')_{ETL}=
\That(qp')G^{(2)N}_c(pq;p'q')_{ETL}} \nonumber \\
&&=\That(qq')G^{(2)N}_c(pq;p'q')_{ETL}=G^{(2)}_c(pq;p'q')_{NETL}.
\end{eqnarray}
\zeqn
{}From eqs.(B.3),(B.5) and (B.12a,b) we have proven that
\begin{equation}
\label{b13}
[G^{(2)}_c]_{any-ETL}=[\That(\alpha)\That(\beta)..
\That(\gamma)G^{(2)N}_c]_{ETL}=[G^{(2)}_c]_{NETL}=i^2 C_2.
\end{equation}

 Now we turn to the general case and carry out the proof by induction.
Assume that eq.(3.4) is valid for $2\leq n \leq m-1 $, i.e.
\begin{equation}
\label{b14}
[i^nG^{(n)}_c]_{any-ETL}=[i^nG^{(n)}_c]_{NETL}=
(-1)^n C_n,(2\leq n \leq m-1).
\end{equation}
We shall prove that it is valid for $2\leq n \leq m$. Eq.(B.14) implies
that
\begin{equation}
\label{b15}
[\That(pq')i^nG^{(n)N}_c]_{ETL}=[i^nG^{(n)}_c]_{NETL}=
(-1)^nC_n, \ (2\leq n \leq m-1).
\end{equation}
Since the cluster expansion (2.14) is a recursive relation, we can
rewrite it as
\begin{equation}
\label{b16}
G^{(m)}_c =G^{(m)}-\A p\S p\sum_{k+l+..+r..+s+t=m}[G^{(k)}_cG^{(l)}_c..
G^{(r)}_c..G^{(s)}_cG^{(t)}_c].
\end{equation}
{}From eq.(B.16) we have
\begin{eqnarray}
\label{b17}
\lefteqn{[(-i)^m\That(pq')G^{(m)N}_c]_{ETL}=
[(-i)^m\That(pq')G^{(m)N}]_{ETL} ,\hspace{2cm} (I)} \nonumber \\
&&-(-i)^m[\That(pq')\A p\S p\sum_{k+l+..+r..+s+t=m}\nonumber\\
&&\cdot (G^{(k)N}_cG^{(l)N}_c..G^{(r)N}_c..G^{(s)N}_cG^{(t)N}_c)]_{ETL},
\hspace{3.8cm} (II).
\end{eqnarray}
The first term on the r.h.s. of eq.(B.17) can be calculated directly,
\begin{eqnarray}
\label{b18}
\lefteqn{(I)=[(-i)^m\That(pq')G^{(m)N}(1..m;1'..m')]_{ETL}}\nonumber\\
&&=[\That(pq')<\psi^{\dagger}(1')..\psi^{\dagger}(q')..
\psi^{\dagger}(m')\psi(m)..\psi(p)..\psi(1)>]_{ETL} \nonumber \\
&&=-<\psi^{\dagger}(1')..\psi(p)..\psi^{\dagger}(m')\psi(m)..
\psi^{\dagger}(q')..\psi(1)>_{ETL} \nonumber \\
&&=<\psi^{\dagger}(1')..\psi^{\dagger}(q')..\psi^{\dagger}(m')
\psi(m)..\psi(p)..\psi(1)>_{ETL} \nonumber \\
&&-\delta^{(3)}(p,q')(-1)^{p+q'}<\psi^{\dagger}(1')..\bar{q'}..
\psi^{\dagger}(m')\psi(m)..\bar{p}..\psi(1)>_{ETL} \nonumber \\
&&-\sum_{j=p+1}^m\delta^{(3)}(j,q')(-1)^{j-q'}<
\psi^{\dagger}(1')..\bar{q'}..\psi^{\dagger}(m')\psi(m)..\bar{j}..
\psi(1)>_{ETL} \nonumber \\
&&-\sum_{i'=q'+1}^{m'}\delta^{(3)}(p,i')(-1)^{i'-p}<\psi^{\dagger}(1')
..\bar{i'}..\psi^\dagger(m')\psi(m)..\bar{p}..\psi(1)>_{ETL}\nonumber\\
&&-\sum_{i'=q'+1}^{m'}\sum_{j=p+1}^m(-1)^{j+i'-p-q'}\delta^{(3)}(p,i')
\delta^{(3)}(j,q')\nonumber\\
&&\cdot <\psi^{\dagger}(1')..\bar{q'}..\bar{i'}..
\psi^{\dagger}(m')\psi(m)..\bar{j}..\bar{p}..\psi(1)>_{ETL} \nonumber\\
&&=(-i)^mG^{(m)}(1..m;1'..m')_{NETL}+(-i)^m \Delta G^{(m)}_{NETL},
\end{eqnarray}
where $\bar{p},\bar{q'},\bar{j}$ and $\bar{i'}$ indicate that they
do not appear in the related green's functions and
$\Delta G^{(m)}_{NETL}$ is
\begin{eqnarray}
\label{b19}
\lefteqn{\Delta G^{(m)}_{NETL}=-i\delta^{(3)}(p,q')(-1)^{p+q'}
G^{(m-1)}(1..\bar{p}..m;1'..\bar{q'}..m')_{NETL} } \nonumber \\
&&-i\sum_{j=p+1}^m\delta^{(3)}(j,q')(-1)^{j-q'}G^{(m-1)}
(1..\bar{j}..m;1'..\bar{q'}..m')_{NETL} \nonumber \\
&&-i\sum_{i'=q'+1}^{m'}\delta^{(3)}(p,i')(-1)^{i'-p}
G^{(m-1)}(1..\bar{p}..m;1'..\bar{i'}..m')_{NETL} \nonumber \\
&&+\sum_{i'=q'+1}^{m'}\sum_{j=p+1}^m\delta^{(3)}(p,i')\delta^{(3)}(j,q')
\nonumber\\
&&\cdot (-1)^{j+i'-p-q'}G^{(m-2)}(1..\bar{p}..\bar{j}..m;1'..\bar{i'}..
\bar{q'}..m')_{NETL}.
\end{eqnarray}
The calculation of the second term of eq.(B.17) is more cumbersome. It reads
\begin{eqnarray}
\label{b20}
\lefteqn{(II)=(-i)^m\That(pq')[(-1)^{p+q'}G^N(p;q')}\nonumber\\
&&\cdot \A p \S p
\sum_{l+..+r..+s+t=m-1}[G^{(l)N}_c..G^{(r)N}_c..G^{(s)N}_c
G^{(t)N}_c](1..\bar{p}..m;1'..\bar{q'}..m')  \nonumber \\
&&+\sum_{j=1}^m(-1)^{j-q'}G^N(j;q')
 \A p\S p\nonumber\\
 &&\cdot \sum_{l+..+r..+s+t=m-1}
[\mbox{terms  where p is not contained in $ G^N$ }]
(1..p..\bar{j}..m;1'..\bar{q'}..m') \nonumber \\
&&+\sum_{i'=1'}^{m'}(-1)^{i'-p}G^N(p;i')\A p \S p\nonumber\\
&&\cdot \sum_{l+..+r..+s+t=m-1}[\mbox{terms where $ q'$ is not contained in
$G^N $}](1..\bar{p}..m;1'.q'.\bar{i'}.m') \nonumber \\
&&-\sum_{i'=1}^{m'}\sum_{j=1}^m(-1)^{j+i'-p-q'}
G^N(j;q')G^N(p;i')\A p\S p \nonumber\\
&&\cdot \sum_{..r+..+s+t=m-2}
[..G^{(r)N}_c..G^{(s)N}_cG^{(t)N}_c]
(1..\bar{p}..\bar{j}..m;1'..\bar{q'}..\bar{i'}..m')
\nonumber \\
&&+ \mbox{other terms where $p,q'$ are not contained in $G^N$}
]_{ETL}.
\end{eqnarray}
Since $\That(pq')$ causes anti-equal time limit, one should consider its
effect term by term. By using eqs.(B.6) and (B.15), we have
\aeqn
\begin{equation}
\label{b21}
\That(pq')G^N(p;q')_{ETL}=G(p;q')_{NETL}-i\delta^{(3)}(p,q'),
\end{equation}
\neqn
\begin{eqnarray}
\That(pq')G^N(j;q')_{ETL}&=&G(j;q')_{NETL},(j\leq p), \nonumber \\
                         &=&G(j;q')_{NETL}-i\delta^{(3)}(j,q'),(j>p),
\end{eqnarray}
\neqn
\begin{eqnarray}
\That(pq')G^N(p;i')_{ETL}&=&G(p;i')_{NETL},(i'\leq q'), \nonumber \\
                         &=&G(p;i')-i\delta^{(3)}(p;i'),(i'>q'),
\end{eqnarray}
\neqn
\begin{equation}
\That(pq')G^{(l)N}_c(...p...;.. q'..)_{ETL}=[G^{(l)}_c]_{NETL},
(2\leq l\leq m-1).
\end{equation}
\zeqn
Inserting the above results in eq.(B.20),we have
\begin{eqnarray}
\label{b22}
\lefteqn{(II)=(-i)^m[(-1)^{p+q'}G(p;q')\A p\S p\sum[G^{(l)}_c..G^{(r)}_c
..G^{(s)}_cG^{(t)}_c]^{(m-1)}(1..\bar{p}.m;1'..\bar{q'}..m')}\nonumber\\
&&+\sum_j(-1)^{j-q'}G(j;q')\A p\S p\sum[G^{(l)}_c..G^{(r)}_c..G^{(s)}_c
G^{(t)}_c]^{(m-1)} (1..\bar{j}..m;1'..\bar{q'}..m') \nonumber \\
&&+\sum_{i'}(-1)^{i'-p}G(p;i')\A p\S p\sum [G^{(l)}_c..G^{(r)}_c..
G^{(s)}_cG^{(t)}_c]^{(m-1)}(1..\bar{p}..m;1'..\bar{i'}..m')\nonumber\\
&&+\sum_{i'}\sum_j(-1)^{j+i'-p-q'}G(j;q')G(p;i')\A p\S p \nonumber\\
&&\cdot \sum[..G^{(r)}_c..G^{(s)}_cG^{(t)}_c]^{(m-2)}(1..\bar{p}..\bar{j}
..m;1'..\bar{q'}..\bar{i'}..m') \nonumber\\
&&+\mbox{other terms where $p,q'$ are not contained in $G$}\nonumber\\
&&(\mbox{the above terms are unlinked terms of $G^{(m)}_{NETL}$})\nonumber\\
&&-i\delta^{(3)}(p;q')(-1)^{p+q'}\A p\S p\sum[G^{(l)}_c..G^{(r)}_c..
G^{(s)}_cG^{(t)}_c]^{(m-1)}(1..\bar{p}..m;1'..\bar{q'}..m')\nonumber\\
&&-i\sum_{j=p+1}^m(-1)^{j-q'}\delta^{(3)}(j;q')(\A p\S
p\sum[G^{(l)}_c..G^{(r)}_c..G^{(s)}_cG^{(t)}_c]^{(m-1)}
(1..\bar{j}..m;1'..\bar{q'}..m') \nonumber \\
&&+\sum_{i'=1}^{m'}(-1)^{i'-p}G(p;i')\A p\S p\sum[..G^{(r)}_c..G^{(s)}_c
G^{(t)}_c]^{(m-2)}(1..\bar{p}..\bar{j}..m;1'..\bar{q'}..\bar{i'}..m'))
\nonumber \\
&&-i\sum_{i'=q'+1}^{m'}(-1)^{i'-p}\delta^{(3)}(p,i')(\A p\S p
\sum[G^{(l)}_c..G^{(r)}c..G^{(s)}_cG^{(t)}_c]^{(m-1)}(1..\bar{p}..m;
1'..\bar{i'}..m') \nonumber \\
&&+\sum_{j=1}^m(-1)^{j-q'}G(j;q')\A p\S p\sum[..G^{(r)}_c..G^{(s)}_c
G^{(t)}_c]^{(m-2)}(1..\bar{p}..\bar{j}..m;1'..\bar{q'}..\bar{i'}..m'))
\nonumber \\
&&-i\sum_{i'=q'+1}^{m'}\sum_{j=p+1}^m(-1)^{j+i'-p-q'}
\delta^{(3)}(p,i')\delta^{(3)}(j,q')\A p\S p \nonumber\\
&&\cdot \sum[..G^{(r)}_c..
G^{(s)}_cG^{(t)}_c]^{(m-2)}(1..\bar{p}..
\bar{j}..m;1'..\bar{q'}..\bar{i'}..m') \nonumber \\
&&(\mbox{the above terms amount to $\Delta G^{(m)}_{NETL}$})]_{NETL}
\nonumber\\
&&=(-i)^m[G^{(m)}(1..m;1'..m')-G^{(m)}_c(1..m;1'..m')]_{NETL}+
(-i)^m\Delta G^{(m)}_{NETL}.
\end{eqnarray}
Comparing eqs.(B.18) and (B.22), we obtain
\begin{equation}
\label{b23}
[\That(pq')G^{(m)N}_c]_{ETL}=[G^{(m)}_c]_{NETL}=(i)^mC_m
\end{equation}
Considering eqs.(B.3),(B.5) and (B.23), we succeed in proving
\begin{equation}
\label{b24}
[G^{(m)}_c]_{any-ETL}=[\That(\alpha)\That(\beta..\That(\gamma)
G^{(m)N}_c]_{ETL}=[G^{(m)}_c]_{NETL}=i^mC_m.
\end{equation}

It is worth to mention that the above proof is equivalent to proving the
consistency of two different  definitions for $G^{(n)}_c$. The first
definition is given by the  cluster expansion (2.14).
The second one can be given as follows: (a) First transform the
T-product to the N-product by Wick's theorem; (b)  then apply the
cluster expansion to the averages of N-products (the normal green's
function  $G^{(n)N}$); we denote the correlation green's function
according to the latter definition by  $G^{(n)N}_c$.
This appendix shows that all the contractions in Wick's expansion can
be absorbed by the one-body normal green's function $G^N$. After
absorption of the contraction, $G^N$ turns out to be $G$, and the second
kind of expansion reduces to the first kind. Thus $G_c^{(n)N}$ and
$G_c^{(n)}$ are equivalent.

\newpage

\section*{Figure captions}
\noindent
{\bf Fig. 1:} Numerical results for $tr \ C_2$ (dashed line)
and $tr \ C_3$ (solid line) within the Fermi-gas model as a function of the
temperature T for a system of 40 particles.

\vspace{1cm}
\noindent
{\bf Fig. 2:} Numerical results for $tr \ C_2$ (dashed line),
$tr \ C_3$ (solid line) and $tr \ C_4$ (dotted line)
for $^{40}Ca$ at 2 MeV temperature
as a function of time within the TDDM approach \cite{r29,r35}.

\end{document}